\documentclass[12pt]{article}
\usepackage{amsfonts}
\usepackage{amssymb}
\usepackage{color}
\usepackage{epic}
\usepackage{graphics}

\def\lddots{\mathinner{\mkern1mu\raise1pt\hbox{.}\mkern2mu
\raise4pt\hbox{.}\mkern2mu\raise7pt\vbox{\kern7pt\hbox{.}}\mkern1mu}}
\makeatletter

\def\numberbysection{\@addtoreset{equation}{section}
\def\theequation{\thesection.\arabic{equation}}}
\makeatother
\numberbysection

\newcommand{\finproof}{{\hfill \rule{5pt}{5pt}}}

\newcommand{\be}{\begin{eqnarray}}
\newcommand{\ee}{\end{eqnarray}}
\newcommand{\non}{\nonumber}

\begin{document}

\begin{titlepage}
\vskip 0.4cm \strut\hfill \vskip 0.8cm
\begin{center}


{\bf {\Large Murphy elements from the double-row transfer matrix}}

\vspace{10mm}


{\large {\bf Anastasia
Doikou}\footnote{adoikou@upatras.gr}}

\vspace{10mm}

{\small University of Patras, Department of Engineering
Sciences,\\ GR-26500 Patras, Greece}

\end{center}

\vspace{30mm}

\begin{abstract}

We consider the double-row (open) transfer matrix constructed from generic tensor-type representations of various types of Hecke algebras. For different choices of boundary conditions for the relevant integrable lattice model we express the double-row transfer matrix solely in terms of generators of the corresponding Hecke algebra (tensor-type realizations). We then expand the open transfer matrix and extract the associated Murphy elements from the first/last terms of the expansion. Suitable combinations of the Murphy elements as has been shown commute with the corresponding Hecke algebra.

\end{abstract}

\vfill \baselineskip=16pt

\end{titlepage}

\section{Introduction}

There has been much activity lately associated to algebraic structures underlying integrable lattice models. On the one hand there is an immediate connection between these models and realizations of the braid group \cite{jimbo0}--\cite{isog}, given that spin chain models may be constructed as tensorial representations of quotients of the braid group called Hecke algebras. On the other hand integrable lattice models provide perhaps the most natural framework for the study of quantum groups \cite{kure, tak}. The symmetry algebras underlying these models may be seen as deformations of the usual Lie algebras \cite{jimbo0, drinf}, and their defining relations emanate directly from the fundamental relations ruling such models, that is the Yang-Baxter \cite{baxter} and reflection equations \cite{cherednik}.

Several studies have been devoted on the uncovering of the symmetries of open spin chain models as well as on connecting the associated Hecke algebras with the underlying quantum group symmetries, and in most cases it turns out that the exact symmetries --quantum algebras-- commute with the Hecke algebra (see e.g. \cite{pasquiersaleur, doikou0, doikouhecke}). In the spin chain context the transfer matrices may be usually expressed in terms of the quantum algebra elements in a universal manner, i.e. independent of the choice of representation of the quantum algebra. However, such generic expressions in terms of Hecke algebra elements are missing, with the exception of generic formulas of integrable Hamiltonians (see e.g. \cite{doikoumartin, doikouhecke} for computational details).
In the present investigation we provide generic expressions, of double-row transfer matrices \cite{sklyanin} in terms of generators of Hecke algebras (tensor type representations).  It is worth noting that such generic expressions starting from Sklyanin's transfer matrix \cite{sklyanin} offer an immediate link between spin chain like systems and other integrable lattice models such as Potts models and in general face type models \cite{baxterfor, martinpots, pearce}. Having such expressions at our disposal we are then able to extract from the double-row transfer matrix the so-called Murphy elements, which commute with the Hecke algebras (see \cite{degi2, fragra} and references therein).

The outline of this paper is as follows. In the next section we give basic definitions regarding the $A,\ B$ and $C$-type Hecke algebras. We also define the Murphy elements associated to each one of the aforementioned Hecke algebras. In section 3 starting from the double-row transfer matrix \cite{sklyanin}  we end up with generic formulas expressed in terms of generators of Hecke algebras (tensor representations). We finally prove that the Murphy elements are directly obtained from suitable double-row transfer matrices of varying dimension. In the last section we briefly discuss the findings of this study, and also propose possible directions for future investigations.

\section{Hecke algebras: definitions}

We shall review in this section basic definitions regarding various types of Hecke algebras, and the associated Murphy elements (see also \cite{fragra}--\cite{hecke5}).
\\
\\
{\underline {{\bf  Definition 2.1.}} {\it The $A$-type Hecke algebra ${\cal H}_{N}(q)$ is defined by the generators $g_{l}$, $l=1,\ldots ,N-1$ satisfying the following relations:}
\be && g_{l}\ g_{l+1}\ g_{l} = g_{l+1}\ g_{l}\ g_{l+1}, \label{br1} \\ && \Big [ g_{l},\ g_{m} \Big ]  =0, ~~~|l-m| >1 \label{br01}\\ &&(g_{l}-q)\ (g_{l} +q^{-1}) =0. \label{braid} \ee
\\
{\underline {{\bf Definition 2.2.}} {\it The $B$-type Hecke algebra ${\cal B}_{N}(q,Q_0)$ is defined by generators $g_{l}$, $l \in \{ 1, \ldots , N-1 \}$, satisfying the Hecke relations (\ref{br1})-(\ref{braid}) and $g_{0}$ obeying:
\be && g_{1}\ g_{0}\ g_{1}\ g_{0} = g_{0}\ g_{1}\ g_{0}\ g_{1}, \label{br2} \\ && \Big [g_{0},\ g_{l} \Big] =0, ~~l>1 \label{br02} \\ && (g_0 - Q_0)\ (g_0 + Q_0^{-1})=0. \label{braid2} \ee}\\
The algebra above is apparently an extension of the Hecke algebra defined in (\ref{braid}). Also the $B$-type Hecke algebra is a quotient of the affine Hecke algebra, which is defined by generators
$g_i,\ g_0$ that satisfy (\ref{br1})-(\ref{br02}).
\\
\\
{\underline {{\bf Definition 2.3.}} {\it  The $C$-type Hecke algebra ${\cal C}_{N}(q,Q_0,Q_N)$, is defined by the generators  $g_{l}$, $l \in \{ 1, \ldots , N-1 \}$, $g_0$ satisfying (\ref{br1})-(\ref{braid2}) and an extra generator $g_{N}$, obeying} \be g_N\ g_{N-1}\ g_N\ g_{N-1} =g_{N-1}\ g_N\ g_{N-1}\ g_N  \\ \Big [ g_{N},\ g_i \Big ], ~~~~~0\leq i \leq N-2 \\ (g_N - Q_N)\ (g_N + Q_N^{-1})=0. \label{quotient} \ee
\\
There is also a quotient of the $C$-type Hecke algebra called the two boundary Temperley-Lieb algebra \cite{tl}--\cite{martinwood}, \cite{levymartin, doikoumartin} with a typical representation being the boundary XXZ model.
\\
\\
{\underline {{\bf  Definition 2.4.}} {\it The two boundary Temperley-Lieb algebra is defined by generators satisfying (\ref{br1})-(\ref{quotient}). In addition to the latter relations some extra equations are also satisfied. Let $e_i = g_i -q,\ e_0 = g_0 -Q_0,\  e_N = g_N -Q_N$ then:} \be && e_i\ e_{i\pm 1}\ e_i = e_i, ~~~~~2 \leq i \leq N-1 \\ && e_1\ e_0\ e_1 = \kappa_-\ e_1 \\ && e_{N-1}\ e_{N}\ e_{N-1} = \kappa_+\ e_{N-1}. \label{TL}  \ee
\\ It is worth mentioning that by removing the third of the above equations we obtain the boundary Temperley-Lieb (blob) \cite{pmartin} algebra, and by removing the second equation as well we end up with the usual Temperley-Lieb algebra \cite{tl}.
\\
\\
{\underline {{\bf Definition 2.5.}} {\it Define also the:}\\
{\it $A$-type Murphy elements} \be J_1^{(A)} &=& g_1^2 \non\\ J_i^{(A)} &=& g_i\ J_{i-1}^{(A)}\ g_i, ~~~~~2 \leq i \leq N-1 \label{amurphy} \ee
{\it $B$-type Murphy elements} \be J_0^{(B)} &=& g_0 \non\\ J_i^{(B)} &=& g_i\ J_{i-1}^{(A)}\ g_i, ~~~~~1
\leq i \leq N-1  \label{bmurphy} \ee
{\it $C$-type Murphy elements}
\be J_0^{(C)} &=& g_1^{-1}\ g_2^{-1} \ldots g_{N-1}^{-1}\ g_N\  g_{N-1} \ldots g_2\ g_1\ g_0 \non\\ J_i^{(C)} &=& g_i\ J_{i-1}^{(C)}\ g_i, ~~~~~1
\leq i \leq N-1.  \label{cmurphy} \ee
\\
It was shown that Murphy elements are pairwise commuting, and symmetric polynomials in $\{J_i^{(A,\ B )} \}$ commute with the $A,\ B$ Hecke algebras respectively (see \cite{degi2, fragra} and references therein,
see also \cite{tysse}). Moreover, symmetric polynomials in $\{ J_i^{(C)},\  (J_i^{(C)})^{-1}\}$ are central in $C$-type Hecke algebras. In the next section we shall show that the Murphy elements defined above arise naturally from certain hierarchies of open transfer matrices.

Let us point out that the $B$-type Murphy elements may be thought of as representations of the so-called $B$-type Artin braid group ${\mathfrak B}_N$ defined by generators $g_i, \ g_0$ and relations (\ref{br1}), (\ref{br01}), (\ref{br2}), (\ref{br02}) --it is evident that the $B$-type Hecke algebra is a quotient of the Artin group ${\mathfrak B}_N$. Such representations are known as the `auxiliary string' representations $\sigma_l: {\mathfrak B}_N \hookrightarrow {\mathfrak B}_{N+l}$ \be \sigma_l(g_0) &=& g_l\ g_{l-1} \ldots g_1\ g_0\ g_1\ \ldots g_{l-1}\ g_l \non\\ \sigma_l(g_i) &=& g_{i+l}  \ee and have been extensively discussed for instance in \cite{doikoumartin, martinwood}. The auxiliary spin representation gives rise to `dynamical' boundary conditions providing extra boundary degrees of freedom (see also relevant discussion in \cite{doikoumartin}).

It will be useful for the following to consider tensor type representations of the Hecke algebra; let $\pi: {\cal C}_N(q, Q_0, Q_N) \hookrightarrow \mbox{End}({\mathbb V}^{\otimes N})$ such that
\be \pi(g_i) &=& {\mathbb I} \otimes {\mathbb I} \ldots \otimes \underbrace{{\mathrm g}}_{\mbox{ $i,\ i+1$}} \otimes \ldots \otimes {\mathbb I} \non\\ \pi(g_0) &=& \underbrace{{\mathrm g}_0}_{\mbox{ $1$}} \otimes\ {\mathbb I} \ldots \otimes {\mathbb I}  \non\\ \pi(g_N) &=& {\mathbb I} \otimes {\mathbb I}  \ldots  {\mathbb I}  \otimes  \underbrace{{\mathrm g}_N}_{\mbox{ $N$}}. \label{rep} \ee
It is clear that the absence of the extra generators $g_N,\ g_0$ leads to representations of the $B$ and $A$ type Hecke algebras.

For instance in the representation of the $C$-type Hecke algebra for the ${\cal U}_q(\widehat{gl_{\cal N}})$ series (${\mathbb V} \equiv {\mathbb C}^{{\cal N}}$) (see also \cite{jimbo, doikouhecke} and \cite{abadrios}) we define:
\be && {\mathrm g} = q {\mathbb I} + \sum_{a \neq b} \Big (e_{ab} \otimes e_{ab} - q^{sgn(a-b)} e_{aa} \otimes e_{bb} \Big ) \non\\ && {\mathrm g}_0 = -Q_0^{-1}e_{11} - Q_0e_{{\cal N}{\cal N}} +x_0^+e_{1{\cal N}} + x^-_0 e_{{\cal N}1} +Q_0{\mathbb I} \non\\ && {\mathrm g}_N = -Q_N e_{11} - Q_N^{-1} e_{{\cal N}{\cal N}} +x_N^+ e_{1{\cal N}} + x^-_N e_{{\cal N}1} +Q_N{\mathbb I} \label{matrices} \ee where ($e_{ij})_{kl} = \delta_{ik} \delta_{jl}$. For ${\cal N} =2$ in particular we recover the well known XXZ representation of the two-boundary Temperley-Lieb algebra.

\section{Murphy elements from open transfer matrices}

Having introduced the basic algebraic setting we are now in a position to extract the above defined Murphy elements from the double-row transfer matrix \cite{sklyanin}. Particular choice of boundary conditions entails Murphy elements associated to the three different types of Hecke algebras defined in the previous section.

Introduce now the Yang-Baxter and reflection equations.
The Yang--Baxter equation is given by \cite{baxter}: \be \check R_{12}(\lambda_{1} -\lambda_{2})\ \check R_{23}(\lambda_{1})\ \check R_{12}(\lambda_{2}) =\check R_{23}(\lambda_{2})\ \check R_{12}(\lambda_{1})\ \check R_{23}(\lambda_{1}-\lambda_{2}) \label{ybe2} \ee acting on ${\mathbb V}^{\otimes 3}$, and as usual $~\check R_{12} = \check R \otimes {\mathbb I}, ~~\check R_{23} = {\mathbb I} \otimes \check R~$.
The reflection equation is also defined as \cite{cherednik} \be \check R_{12}(\lambda_{1} -\lambda_{2})\ K_{1}(\lambda_{1})\ \check R_{12}(\lambda_{1} +\lambda_{2})\ K_{1}(\lambda_{2})=K_{1}(\lambda_{2})\ \check R_{12}(\lambda_{1} +\lambda_{2})\ K_{1}(\lambda_{1})\ \check R_{12}(\lambda_{1} -\lambda_{2}) \label{re2} \ee acting on ${\mathbb V}^{\otimes 2} $, and as customary $~K_{1} = K \otimes {\mathbb I}$, $~ K_{2} = {\mathbb I} \otimes  K$. Notice the structural similarity between the Yang-Baxter and reflection equation and the Hecke algebras above, which suggests that representations of ${\cal B}_{N}(q,Q)$  should provide candidate solutions of the Yang-Baxter and reflection equations. To construct a spin chain like system with two non-trivial boundaries we shall need to consider one more reflection equation associated to the other end of the $N$ site spin chain, i.e.
\be && \check R_{N-1\ N}(\lambda_{1} -\lambda_{2})\ \bar K_{N}(\lambda_{1})\ \check R_{N-1\ N}(\lambda_{1} +\lambda_{2})\ \bar K_{N}(\lambda_{2})  \non\\ && = \bar K_{N}(\lambda_{2})\ \check R_{N-1\ N}(\lambda_{1} +\lambda_{2})\ \bar K_{N}(\lambda_{1})\ \check R_{N-1\ N}(\lambda_{1} -\lambda_{2}). \label{re3} \ee

Consider solutions of the Yang-Baxter and reflection equations in terms of the generators of the $C$-type Hecke algebra:
$g_0,\ g_1,\ \ldots g_{N-1},\ g_N$,
\be
\check R_{i\ i+1}(\lambda)  &=& e^{\lambda} g_i - e^{-\lambda} g_i^{-1}, ~~~~~~~i \in \{1, \ldots ,N-1 \}
\non\\ K_1(\lambda) &=&  e^{2 \lambda} g_0 + c_{-} -  e^{-2\lambda} g_0^{-1} \non\\ \bar K_N(\lambda) &=& e^{2 \lambda} g_N + c_{+} - e^{-2\lambda} g_N^{-1}
\label{sol}
\ee the boundary parameters are incorporated in $g_0,\ g_N$.
Note also that $\check R$ and $K$ matrices are unitary i.e. \be \check R_{12}(\lambda)\ \check R_{12} (-\lambda) \propto {\mathbb I}, ~~~~~K_1(\lambda)\ K_1(-\lambda) \propto {\mathbb I}, ~~~~~\bar K_N(\lambda)\ \bar K_N(-\lambda) \propto {\mathbb I}. \ee
Recall that $R_{ij} = {\cal P}_{ij}\ \check R_{ij}$, where ${\cal P}$ is the permutation operator, and the $R$ matrix in general satisfies \be && R_{12}^{t_1}(\lambda)\ M_1 \ R_{12}(-\lambda - 2 \rho)^{t_2}\ M_1^{-1} \propto {\mathbb I} \label{property}\\ && \mbox{with} ~~~~~~~\Big [ M_1\ M_2,\ R_{12}(\lambda) \Big ] =0, ~~~~~M^t =M, \ee $\rho$ is the crossing parameter, and for instance in the ${\cal U}_q(\widehat{gl_{\cal N}})$ case $\rho = {{\cal N} \over 2}$.
The latter property
(\ref{property}) together with unitarity and the use of reflection equation are essential in proving the integrability of an open integrable lattice model \cite{sklyanin}.
Note that $M$ is modified according to the choice of representation (see
\cite{doikoumartin, doikouhecke}). For instance $M$ for the ${\cal U}_q(\widehat{gl_{\cal N}})$ series \cite{jimbo} is given by the diagonal ${\cal N} \times {\cal N}$ matrix  (see also \cite{doikouhecke}): \be M =  q^{{\cal N}-2j+1} \delta_{ij}. \ee

Henceforth we shall focus on tensorial representations of Hecke algebras of the type (\ref{rep}), although still we do not choose any particular representation --i.e. the form of ${\mathrm g},\ {\mathrm g}_0,\ {\mathrm g}_N$ in (\ref{rep}) is not specified, is kept generic--, so the subsequent propositions and proofs are quite generic. Also for simplicity we set $\pi(g_i) \equiv g_i$.

With the above general setting at our disposal we may now show the following propositions:
\\
\\
{\underline {\bf Proposition 1}}: {\it Tensor representations of the Murphy elements associated to the $B$-type Hecke algebra are obtained from the hierarchy of double-row transfer matrices with one non-trivial boundary}:
\be
&& t^{(n)}(\lambda) = tr_0 \Big \{M_0\ R_{0n}(\lambda +\lambda_0) R_{0n-1}(\lambda)
\ldots R_{01}(\lambda)\ K_0(\lambda)\ R_{10}(\lambda) \ldots R_{n0}(\lambda -\lambda_0)\Big \}  \non\\
&& 1 \leq n \leq N, ~~~~t^{(n)}(\lambda) \in \mbox{End}({\mathbb V}^{\otimes n})
\label{t0} \ee {\it provided that}: \be tr_0 \{M_0\ \check R_{n0}(2\lambda_0) \} \propto {\mathbb I}. \label{req1} \ee \\
{\underline {\it Proof}}: Notice the presence of the inhomogeneity $\lambda_0$ at the $n^{th}$ site. In general we could have set
inhomogeneities everywhere, but for our purposes here it is sufficient to consider only $\lambda_0$. Consider also that $\lambda = \lambda_0$, with $~\lambda_0$ being a free parameter, then the transfer matrix becomes
\be
t^{(n)}(\lambda_0) = tr_0 \{M_0\ \check R_{n0}(2\lambda_0) \}\ \check R_{n-1\ n}(\lambda_0) \ldots \check R_{12}(\lambda_0) K_1(\lambda_0) \check R_{12}(\lambda_0)\
\ldots  \check R_{n-1\ n}(\lambda_0).\non\\ \label{tr}
\ee
Although (\ref{req1}) is a requirement in our proof it is relatively easy to show for instance that for the  ${\cal U}_q(\widehat{gl_{\cal N}})$ series \cite{jimbo}, (\ref{req1}) is valid. Taking into account (\ref{req1}) we have
\be
t^{(n)}(\lambda_0) \propto \check R_{n-1\ n}(\lambda_0)
\check R_{n-2\ n-1}(\lambda_0) \ldots \check R_{12}(\lambda_0) K_1(\lambda_0) \check
R_{12}(\lambda_0) \ldots  \check R_{n-1\ n}(\lambda_0)
\ee
and bearing in mind (\ref{sol}) we conclude:
\be
t^{(n)}(\lambda_0) & \propto & (g_{n-1} - e^{-2 \lambda_0}g_{n-1}^{-1}) \ldots
(g_1 - e^{-2 \lambda_0} g_1^{-1}) (g_0 + e^{-2\lambda_0} c_- -e^{-4 \lambda_0} g_0^{-1})\non\\ &\times&
(g_1 - e^{-2 \lambda_0} g_1^{-1}) \ldots (g_{n-1} - e^{-2 \lambda_0}g_{n-1}^{-1}). \label{tr2}
\ee
The open spin chain transfer matrix is eventually expressed solely in terms of the $B$-type Hecke algebra generators. And if we expand the transfer matrix in powers of $e^{-2\lambda_0}$ we end up with:
\be
t^{(n)}(\lambda_0) &\propto &
g_{n-1}\ g_{n-2} \ldots g_1\ g_0\ g_1 \ldots g_{n-2}\ g_{n-1} +\ldots \non\\
&-& e^{-4 n \lambda_0} g^{-1}_{n-1}\ g^{-1}_{n-2} \ldots g^{-1}_1\ g^{-1}_0\ g^{-1}_1 \ldots g^{-1}_{n-2}\ g^{-1}_{n-1}. \label{exp} \ee
The first and last term of the expansion above are clearly the Murphy element $J_{n-1}^{(B)}$ and its opposite respectively. \finproof\\
\\
{\underline {{\bf Corollary}}: {\it Tensor representations of the $A$-type Murphy elements are obtained from the hierarchy of transfer matrices (\ref{t0}) for
$K^- \propto {\mathbb I}$.}\\
{\underline {\it Proof}}: the proof of this statement is straightforward; in this case evidently $g_0 \propto {\mathbb I}$. \finproof\\

Consider now the matrices $K^-(\lambda)$ and $K^+ = K^t(-\lambda -i\rho)$ where $K^-,\ K$ are solutions of the reflection equation (\ref{re2}). Consider also the dynamical type solutions
of the reflection equations (\ref{re2}) and (\ref{re3}) respectively:
\be {\mathbb K}_0^-(\lambda) &=&
R_{0N}(\lambda +N\delta) \ldots R_{02}(\lambda +2 \delta)\ R_{01}(\lambda +\delta)\ K_0^-(\lambda)\ R_{10}(\lambda-\delta) \ldots R_{N0}(\lambda -N\delta)\non\\
 {\mathbb K}_0^+(\lambda) &=&
R_{10}(\tilde \lambda+\delta) R_{02}(\tilde \lambda -2\delta)
\ldots R_{N0}(\tilde \lambda-N\delta)\ K^+_0(\lambda)\ R_{0N}(\tilde \lambda +N\delta) \ldots R_{01}(\tilde \lambda -\delta) \non\\ \tilde \lambda &=& - \lambda - i\rho.
\label{t} \ee Then it can be shown that:
\\
\\
{\underline {{\bf Proposition 2}}: {\it Tensor representations of the Murphy elements $(J_{N-1}^{(C)})^{\pm 1}$, $(J_0^{(C)})^{\pm 1}$ associated to the $C$-type Hecke algebra are obtained from the following open transfer matrices with two non-trivial boundaries}
\be && t^{(-)}(\lambda) = tr_0 \Big \{M_0\ K_0^+(\lambda)\  {\mathbb K}_0^-(\lambda) \Big \}, ~~~~~~ t^{(+)}(\lambda) = tr_0 \Big \{M_0\ {\mathbb K}_0^+(\lambda)\  K_0^-(\lambda) \Big \} \non\\ && t^{(\pm)}(\lambda) \in \mbox{End}({\mathbb V}^{\otimes N}) \label{t1} \ee
{\it provided that}: \be
tr_{0}\{ M_0\ K_0^+(\lambda)\ \check R_{N0}(2\lambda) \} \propto \bar K_N(\lambda), ~~~~tr_{0}\{  K_0^-(\lambda -i\rho)\ M_0\ \check R_{10}(-2\lambda) \} \propto K_1(\lambda).
\label{condition2} \ee
\\
{\underline {\it Proof}}: Notice the main difference with the previous case, $N$ the length of the spin chain is now fixed, whereas previously the length of the chain was variable. The presence of the second non-trivial boundary fixes somehow the length of the chain and this is already evident when defining the $C$-type Murphy elements.

We start with the $t^{(-)}$ matrix, we set $\lambda = N \delta$ then the double-row transfer matrix becomes: \be
t^{(-)}(N\delta) &=& tr_0 \{M_0\ K_0^+(N \delta) \check R_{N0}(2N\delta) \}\non\\ & \times &  \check R_{N-1\ N}((2N-1)\delta)
\ldots \check R_{12}((N+1)\delta) K_1( N \delta)\ \check R_{12}((N-1)\delta)\
\ldots  \check R_{N-1\ N}(\delta) \non\\ \label{tr+}
\ee
It can be explicitly checked, that conditions (\ref{condition2}) are valid for instance for the ${\cal U}_q(\widehat{gl_{\cal N}})$ series. Bearing in mind (\ref{condition2}) we conclude
\be
t^{(-)}(N\delta) & \propto & (g_N + c_{+} e^{-2 N \delta}- e^{-4 N \delta}
g_N^{-1}) (g_{N-1} - e^{-2 (2N-1) \delta}g_{N-1}^{-1}) \ldots (g_1 - e^{-2(N +1)\delta} g_1^{-1})
\non\\ &\times& (g_0 + e^{-2N \delta} c_-
-e^{-4 N \delta} g_0^{-1})
(g_1 - e^{-2 (N-1)\delta} g_1^{-1}) \ldots (g_{N-1} - e^{-2 \delta}g_{N-1}^{-1}). \non\\
\label{tr2+}
\ee In this case the double-row transfer matrix is expressed in terms of $C$-type Hecke algebra generators, and by expanding in powers of $e^{-\delta}$ we get:
\be t^{(-)}(N \delta) & \propto & g_N\ g_{N-1}\ g_{N-2} \ldots g_1\ g_0\ g_1 \ldots g_{N-2}\ g_{N-1} + (\mbox{higher order terms}) \label{exp2}
\ee the first term of the expansion above is the Murphy element $J_{N-1}^{(C)}$.

Similarly for the $t^{(+)}$ matrix we set $-\lambda - i\rho = -\delta$ then:
\be t^{(+)} (-\delta + i\rho) &=& \check R_{12}(-3\delta) \dots \check R_{N-1 N}(-(N+1)\delta) K_N(\delta) \check R_{N-1 N}((N-1)\delta) \ldots \check R_{12}(\delta) \non\\ & \times & tr_0\Big \{\check R_{10}(-2\delta) K_0(\delta - i\rho)M_0 \Big \}. \ee Bearing in mind the expressions of $K$ and $\check R$ matrices in terms of the Hecke algebra generators, and (\ref{condition2}) we may rewrite the $t^{(+)}$ as: \be t^{(+)}(-\delta +i\rho ) &=& (-1)^{N} (g_1^{-1} - e^{-6\delta} g_1) \ldots
(g^{-1}_{N-1} - e^{-2(N+1)\delta}g_{N-1}) (g_N + c_+ e^{-2\delta} - e^{-4\delta} g^{-1}_N) \non\\ &\times & (g_{N-1} - e^{-2(N-1)\delta} g_{N-1}^{-1}) \ldots (g_1 - e^{-2 \delta} g_1^{-1})(g_0 +c_- e^{-2\delta}- e^{-4\delta} g_0^{-1})\ee
and finally by expanding $t^{(+)}$ we conclude
\be
t^{(+)}(N \delta) & \propto & g_1^{-1}\ g_{2}^{-1}  \ldots g_{N-1}^{-1}\ g_N\ g_{N-1} \ldots g_1\ g_0  +(\mbox{higher order terms}). \label{exp3}
\ee
Notice that the zero order term in the expansion (\ref{exp3}) is the element $J_0^{(C)}$.
The opposite Murphy elements $(J_{N-1}^{(C)})^{-1},\ (J_{0}^{(C)})^{-1}$ can be obtained from $t^{(-)},\ t^{(+)}$ at $\lambda =-N \delta$ and $-\lambda -i\rho =\delta$ respectively as the zero order terms in the corresponding expansions. Notice that in this case we are able to extract only the $J_{N-1}^{(C)},\ J_0^{(C)}$ elements and their opposites contrary to the previous case, where all the Murphy elements were extracted from the transfer matrices (\ref{t0}). Some comments on this intricate issue will be presented in the discussion section below, however a more detailed investigation will be pursued elsewhere. It is finally clear from the expressions above that for $g_N \propto {\mathbb I}$ the results of Proposition 1 are recovered. \finproof\\

Assuming the expansion around $\lambda= \delta=0$ one obtains local integrals of motion (we refer the interested reader to \cite{doikoumartin, doikouhecke} for a more detailed discussion). For instance the first derivative of (\ref{tr2}) with respect to $\lambda$ (at $\lambda =0$) gives the well
known Hamiltonians discussed also e.g. in \cite{doikoumartin, doikouhecke}, expressed as a sum of the Hecke elements (better set $g_i = e_i +q,\ g_0 =e_0 +Q_0,\ g_N = e_N + Q_N$). Higher terms in such an expansion provide naturally higher Hamiltonians.

\section{Discussion}

We have been able to extract all $A$ and $B$-type Murphy elements from suitable hierarchies of open transfer matrices (\ref{t0}).
For the moment we have been able to only identify the $C$-type elements $J_{N-1}^{(C)},\ J_0^{(C)}$ and their opposites from the open transfer matrices $t^{(\pm)}$ (\ref{t1}). Note that we mainly focused on a big class of generic tensor type (spin chain like) representations (\ref{rep}) of Hecke algebras, however we did not choose any special representation i.e. we did not consider any particular form for the quantities ${\mathrm g},\ {\mathrm g}_0,\ {\mathrm g}_N$ appearing in (\ref{rep}), hence our results are generic.

We assume that a generic choice of an integrable spin chain with two suitable dynamical ${\mathbb K}^{\pm(n)}$ reflection matrices involving an appropriate sequence of inhomogeneities, would give all the $C$-type Murphy elements.
In other words the procedure described above may be seen as a convenient prescription that provides relatively easily the Murphy elements. The idea however is to search for a more systematic approach to tackle this problem.
Consider a generic transfer matrix of the form
\be  t^{(n)}(\lambda) &=&  tr_0 \Big \{ M_0\ {\mathbb K}_0^{+(n)}(\lambda)\ {\mathbb K}_0^{-(n)}(\lambda) \Big \}\ee where we define \be  {\mathbb K}_0^{+(n)}(\lambda)&=& R_{0n+1}(\tilde \lambda -( n+1)\delta)R_{0n+2}(\tilde \lambda -( n+ 2)\delta) \ldots  R_{0N}(\tilde \lambda -N\delta)  \non\\ &\times& K_0^+(\lambda) R_{N0}(\tilde \lambda +N \delta) \ldots R_{n+10}(\tilde \lambda + (n+1)\delta) \non\\ {\mathbb K}_0^{-(n)}(\lambda)&=& R_{0n}( \lambda +n\delta)R_{0n-1}(\lambda +( n- 1)\delta) \ldots  R_{01}(\lambda +\delta)  \non\\ &\times& K_0^-(\lambda) R_{10}(\lambda -\delta) \ldots R_{n0}(\lambda -n\delta)\ee recall $\tilde \lambda = -\lambda -i\rho$. This generic type of transfer matrices will presumably provide all the Murphy elements of $C$-type defined in (\ref{cmurphy}).
So as in the $B$-type Hecke case we better deal with an hierarchy of open transfer matrices of varying length or more precisely of modified `dynamics' as far as the boundaries are concerned. Of course one has to take special care when choosing the suitable inhomogeneity to expand around as well as when taking the trace over the auxiliary space, given that certain quite complicated identities involving dynamical $K$-matrices are needed. These however are rather technically involved issues and will be left for future investigations

In the case of two non-trivial boundary XXZ chain the Murphy elements, could be expressed in terms of the charges in involution, and as such should be also expressed in terms of the abelian part of the $q$-Onsager algebra derived in \cite{pascal1, pascal}. More precisely the question raised is whether the relevant Murphy
elements can be expressed in terms of the fundamental objects, the so-called boundary non-local charges  (see e.g. \cite{doikou0, pascal1, dema, anchku}), that generate the $q$-Onsager algebra \cite{pascal1, pascal}.

In general for the ${\cal U}_q(\widehat{gl_{\cal N}})$ series the Murphy elements consist an abelian algebra. It has been shown however in \cite{doikouhecke} that there exist a set of centralizers that form a non-abelian algebra --the boundary quantum algebra--, which may be thought of as the analogue of upper/lower Borel subalgebra in ${\cal U}_q(\widehat{gl_{\cal N}})$.
In \cite{doikou0, doikouhecke} the boundary non-local charges (centralizers of the $B$-type Hecke algebra) are extracted from the asymptotics of the tensor representation of the reflection algebra, so it should be possible to see relations among the Murphy elements and boundary non-local charges in the general case. We hope to address these intriguing issues in forthcoming publications.
\\
\\
\noindent{\bf Acknowledgments:} I am indebted to J. de Gier and P. Pearce for illuminating discussions.

\end{document}